\documentclass[aps,prl,reprint,superscriptaddress]{revtex4-2}
\usepackage{graphicx}
\usepackage{amsmath}
\usepackage{amssymb}
\usepackage{natbib}
\usepackage{tabularray}
\usepackage{multirow}
\usepackage[colorlinks=true, linkcolor=blue, citecolor=blue, anchorcolor=blue]{hyperref}

\begin{document}

\title{Manipulation of topology by electric field in breathing kagome lattice}
\author{Yu Xie}
\affiliation{School of Materials Science and Physics, China University of Mining and Technology, Xuzhou 221116, China}
\author{Ke Ji}
\affiliation{Department of Physics, University of Antwerp, Groenenborgerlaan 171, B-2020 Antwerp, Belgium}
\author{Jun He}
\affiliation{School of Materials Science and Physics, China University of Mining and Technology, Xuzhou 221116, China}
\author{Xiaofan Shen}
\affiliation{Department of Physics, University of Antwerp, Groenenborgerlaan 171, B-2020 Antwerp, Belgium}
\author{Dinghui Wang}
\affiliation{School of Materials Science and Physics, China University of Mining and Technology, Xuzhou 221116, China}
\author{Junting Zhang}
\email{juntingzhang@cumt.edu.cn}
\affiliation{School of Materials Science and Physics, China University of Mining and Technology, Xuzhou 221116, China}

\renewcommand{\figurename}{FIG.}
\renewcommand\thefigure{\arabic{figure}}

\begin{abstract} 
Magnetic kagome lattices have attracted much attention because the interplay of band topology with magnetism and electronic correlations give rise to various exotic quantum states. A common structural distortion in the kagome lattice is the breathing mode, which can significantly influence the magnetism and band characteristics. However, the control of breathing mode and the associated topological phenomena remain rarely explored. Here, we demonstrate that the coupling of breathing modes with ferroelectricity, magnetism, and band topology in the $\textit{M}_{3}\textit{X}_{8}$ monolayer system enables electric field manipulation of topological spin structure and electronic states. The breathing mode mainly occurs in materials containing early $4\textit{d}/5\textit{d}$ transition metal elements and can be reversed or even suppressed via ferroelectric switching in low-barrier materials. Importantly, electric field-induced switching of the breathing mode can alter the chirality of the topological spin structure, or trigger a transition from a topological trivial insulator to a Chern insulator. This work paves the way for exploring novel physical phenomena driven by breathing mode in kagome materials.
\end{abstract}
\maketitle

The interplay of non-trivial topology, magnetism, and electron correlations in kagome lattice leads to diverse emerging phenomena and novel quantum states \cite{Ortiz2020,Liang2021,Zhou2017,Kang2019,Yin2022,Wang2023}. For instance, \emph{A}V$_3$Sb$_5$ (\emph{A} = K, Rb,Cs), a representative class of kagome materials, has attracted significant attention for exhibiting unconventional superconductivity, charge and pair density waves, nematic phase, Dirac semimetal, and large anomalous Hall effect in a single material \cite{Ortiz2021,Mielke2022,Nie2022,Chen2021,Li2021,Xu2022,Kang2022,Yang2020}. The combination of intrinsic magnetism and topology in kagome magnets causes a variety of topological electronic states \cite{Xu2015,Ye2018,Tang2011,Liu2022}. In addition, real space topological spin textures, magnetic skyrmions, have been identified in some frustrated kagome magnets \cite{Hou2018,Li2024,Hirschberger2019}.

A common structural distortion in the kagome lattice is the breathing mode \cite{Wang2023}, which can not only modulate electron correlation, magnetic coupling and band structure, but also break the spatial inversion symmetry \cite{Hirschberger2019,Ezawa2018,Bolens2019,Pasco2019}. Therefore, the breathing kagome lattice provides a broad platform for exploring various physical effects related to inversion symmetry breaking \cite{Peng2020,Li2021a,Zhao2024,Hu2024,Xing2024}. Nb$_3$\emph{X}$_8$ (\emph{X} = Cl, Br, I), a representative material family with a breathing kagome lattice, has gained attention as an emerging class of kagome materials for its exotic correlated electronic behaviors, such as topological flat bands \cite{Sun2022,Regmi2022,Zhang2023}, Mott-insulating state \cite{Gao2023,Grytsiuk2024}, and quantum spin liquid state \cite{Schaffer2017,Hu2023}. Due to their weak van der Waals coupling, layered Nb$_3$\emph{X}$_8$ crystals can be easily exfoliated into monolayers \cite{Pasco2019}, making them an ideal system for studying novel physical effects driven by breathing mode.

Breathing mode may provide an additional degree of freedom to manipulate the exotic electronic states and magnetism of kagome materials. However, its control via external fields, particularly electric field, has not yet been demonstrated. In this work, we screen materials with breathing mode in the \emph{M}$_3$\emph{X}$_8$ (\emph{M} involves all transition metal ions and \emph{X} = Cl, Br, I) monolayer system, and then demonstrate that an out-of-plane pulsed electric field can reverse the breathing mode or even suppress it through a ferroelectric-to-paraelectric phase transition. Altering the sign or amplitude of the breathing mode with an electric field enables the chirality of topological spin textures to be switched or induces a topological phase transition, respectively.  

\begin{figure*}
\centering
\includegraphics*[width=0.7\textwidth]{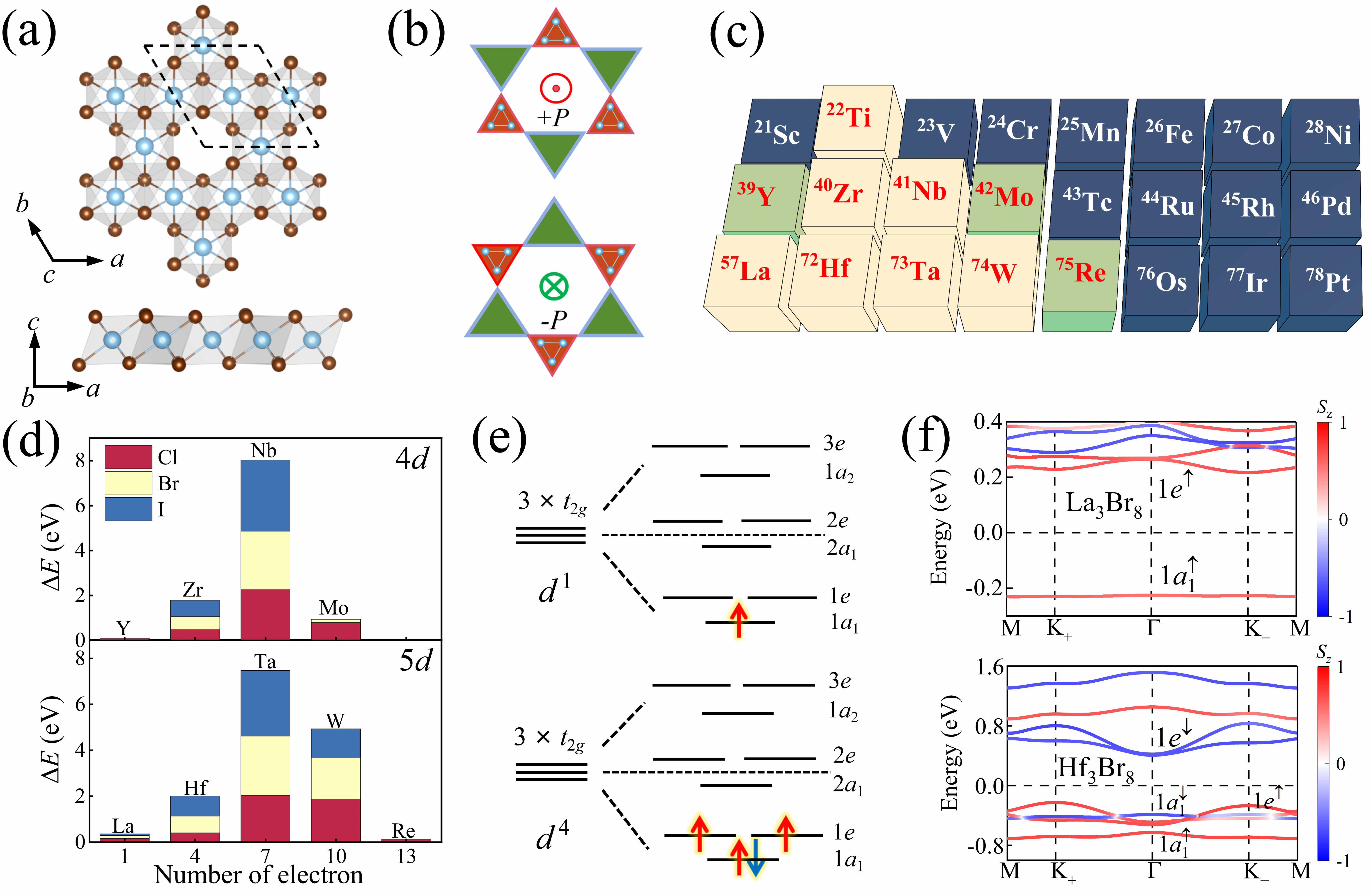}
\caption{\label{Fig1}  (a) Top and side views of the crystal structure of \emph{M}$_3$\emph{X}$_8$ monolayers. (b) Out-of-plane polarization induced by breathing mode and its sign switching. (c) Screened materials with breathing mode in their ground-state structures. Yellow and dark blue fills represent, respectively, the presence and absence of breathing mode in materials. The green fill represents that the breathing mode only occurs in materials containing certain halogen elements. (d) Energy barrier for switching breathing mode. The number on the horizontal axis represents the number of \emph{d} electrons contained in each \emph{M}$_3$ trimer. (e) Schematic illustration of the molecular orbital formed by trimerization and the two configurations of electron occupation. (f) Spin-resolved band structures of La$_3$Br$_8$ and Hf$_3$Br$_8$, corresponding to the two occupation configurations in (e), respectively.}
\end{figure*}

The \emph{M}$_3$\emph{X}$_8$ monolayer consists of edge-sharing octahedra, in which the transition metal ions located at the center of the octahedron form a kagome lattice sandwiched by two sheets of halogen ions [see Fig. \ref{Fig1}(a)]. In the breathing lattice, transition metal ions are trimerized through forming metal-metal bonds, resulting in alternating arrangements of adjacent large and small triangles [see Fig. \ref{Fig1}(b)]. This structural distortion breaks the spatial inversion symmetry, thus allowing the emergence of out-of-plane polarization. Polarization can be reversed by switching the breathing mode, wherein the trimerization shifts to the adjacent large triangle [see Fig. \ref{Fig1}(b)]. At present, Nb$_3$\emph{X}$_8$ is the only experimentally confirmed material family in this system with breathing mode \cite{Pasco2019}. A few other \emph{M}$_3$\emph{X}$_8$ (\emph{M} = Ti, Ta, W) monolayers have been predicted to exhibit breathing mode, where ferroelectricity induced by breathing mode was proposed, but generally with very high energy barriers that prevent the switchability of polarization \cite{Li2021a,Zhao2024,Hu2024,Xing2024}.

We first screened materials with breathing mode throughout the \emph{M}$_3$\emph{X}$_8$ monolayer system. As shown in Fig. \ref{Fig1}(c), the breathing mode is widely present in materials containing early 4\emph{d} and 5\emph{d} transition metal ions, whereas for the 3\emph{d} series, it only occurs in Ti$_3$\emph{X}$_8$ monolayers. This phenomenon is related to the fact that 4\emph{d}/5\emph{d} orbitals are more extended and thus more prone to form metal-metal bonds. All materials with breathing mode in their ground-state structures show dynamic stability except for Re$_3$I$_8$ monolayer (see Supplementary Material Fig. S1 \cite{SM}).

For the 4\emph{d} and 5\emph{d} series, the energy reduction caused by the breathing mode exhibits a non-monotonic trend with increasing atomic number, reaching the maximum at the middle Nb and Ta, respectively [see Fig. \ref{Fig1}(d)]. This trend is attributed to the change in molecular orbital occupation in an isolated \emph{M}$_3$\emph{X}$_{13}$ cluster, consisting of three edge-sharing octahedra. The number of electrons occupying the molecular orbitals is 3\emph{n}-2, where \emph{n} represents the order number of \emph{M} element in the $4d/5d$ series. When the occupancy is less than half full (3\emph{n}-2 $<$ 9), the energy levels of all occupied molecular orbitals are lower than that of the $\emph{t}_{2\emph{g}}$ orbitals, resulting in an increase in energy gain from trimerization with the increase in the number of valence electrons [see Fig. \ref{Fig1}(e)]. When the occupancy exceeds half full, the energy level of the highest occupied molecular orbital is higher than that of the $\emph{t}_{2\emph{g}}$ orbitals. Consequently, the energy gain gradually decreases as the number of valence electrons increases. This trend further explains why the breathing mode only appears in the early part of the transition metal series. 

\begin{figure}
\centering
\includegraphics*[width=0.45\textwidth]{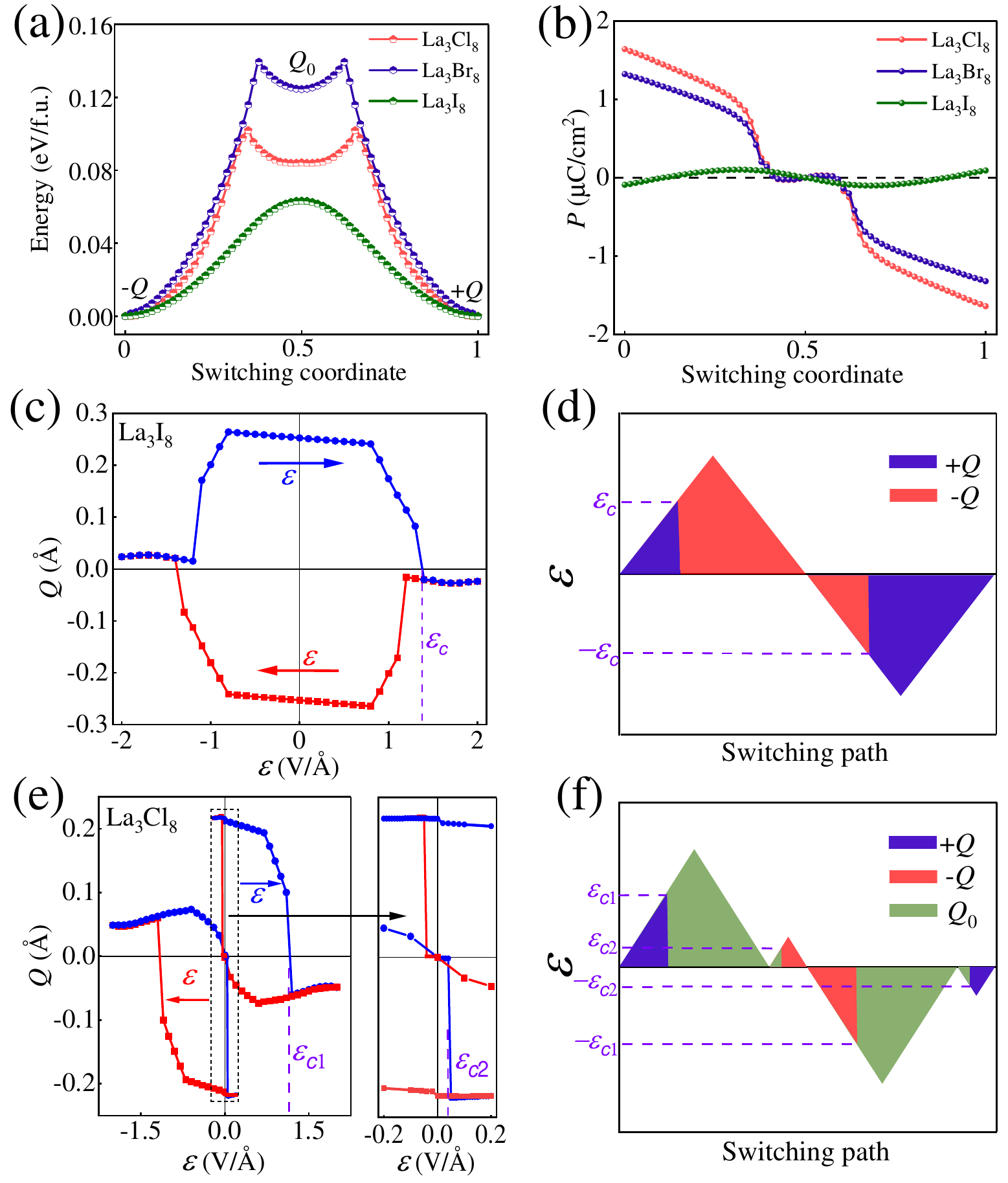}
\caption{\label{Fig2} Changes in (a) energy and (b) polarization of La$_3$\emph{X}$_8$ monolayers along the ferroelectric switching path. Changes in the (c) amplitude \emph{Q} (represented by the cation displacement) and (d) sign of breathing mode in La$_3$I$_8$ monolayer induced by pulsed electric field. Each point represents an optimized structure under a specific electric field, with its initial structure being the equilibrium structure under the previous electric field. The blue and red arrows represent the process of increasing and decreasing the electric field, respectively. (e) Electric field-induced change in the breathing mode of La$_3$Cl$_8$ monolayer. The right inset shows the switching of the breathing mode induced by a small electric field near the paraelectric phase. (f) The applied pulse electric field and resulting sign switching of breathing mode. \emph{Q}$_0$ represents the intermediate state in which the breathing mode disappears upon removal of the electric field.}
\end{figure}

Depending on the parity of \emph{n}, the highest occupied states are the up spin states of the single and doubly degenerate orbitals, respectively [see Fig. \ref{Fig1}(e)], resulting in a total magnetic moment of 1 or 2 $\mu_{B}$ for each cluster. All materials with breathing mode exhibit insulating behavior. The band gap may be opened by crystal field splitting or Coulomb repulsion interaction, corresponding to the charge-transfer type and Mott insulators, respectively [see Fig. \ref{Fig1}(f)]. 

For materials with cations in the middle of the series, the high energy barrier prevents their polarization from being reversed by an electric field. To verify the accessibility of ferroelectric switching, we focus on materials with cations at the beginning of the 5\emph{d} series, La$_3$\emph{X}$_8$ (\emph{X} = Cl, Br, I) monolayers, which exhibit low energy barriers. Interestingly, the La$_3$Cl$_8$ and La$_3$Br$_8$ monolayers exhibit potential well-type energy curves around the paraelectric phase [see Fig. \ref{Fig2}(a)], which show dynamic and thermodynamic stability (Fig. S2 \cite{SM}) . This change in trend can be attributed to the emergence of trimerization caused by enhanced breathing mode, which leads to a metal-insulator transition (Fig. S3 \cite{SM}). The ferroelectric polarization also exhibits a nonmonotonic variation along the switching path, and undergoes a sign change when transitioning toward the paraelectric phase [see Fig. \ref{Fig2}(b)]. The two polar modes contributing to the polarization are both coupled to the breathing mode, but they exhibit different trends (Fig. S4 \cite{SM}), which may be responsible for the complex behavior of ferroelectric polarization.

Figure \ref{Fig2}(c) shows the evolution of the breathing mode in La$_3$I$_8$ monolayer under one electric field cycle. The amplitude of the breathing mode changes slightly initially and then decreases rapidly as the electric field increases. It undergoes a sign reversal when the electric field reaches a critical value $\varepsilon_c$, and remains roughly constant with further increase in the electric field. When the electric field decreases below the critical value, the breathing mode gradually transitions to the state opposite to its initial one. Then applying an electric field in the opposite direction causes the breathing mode to undergo a similar transition process and eventually return to its initial state. Figure \ref{Fig2}(d) depicts the applied electric field pulse and the corresponding switch in the sign of the breathing mode. Other modes also undergo similar transitions, among which the polar mode of anions at general positions shows a strong positive correlation with the electric field, suggesting it as the primary driving force for electric field-induced structural transformation (Fig. S5 \cite{SM}). The electric field control of the breathing mode arises from the linear coupling effect between the breathing mode and the polar displacement mode (Fig. S6 \cite{SM}), which is allowed to exist in the Landau free energy function since they belong to the same irreducible representation $\Gamma_2^-$.

Interestingly, the breathing mode of La$_3$Cl$_8$ monolayer exhibits a more complex behavior under the action of a cyclic electric field [see Fig. \ref{Fig2}(e)]. When the electric field exceeds a critical value, the amplitude of the breathing mode decreases sharply and undergoes a sign change, similar to the behavior of La$_3$I$_8$ monolayer. However, the difference is that as the electric field decreases, the breathing mode gradually disappears instead of transitioning to a state opposite to the initial one. Then when a small positive electric field is applied, the breathing mode changes dramatically to the opposite state from the initial one [see the inset of Fig. \ref{Fig2}(e)]. Applying an electric field in the opposite direction induces a similar transition, returning the breathing mode from the opposite state to its initial state. Therefore, ferroelectric switching is accomplished in two steps, that is, first switching to the paraelectric phase by applying a larger pulse electric field, and then transitioning to the opposite polarization state via a smaller pulse electric field [see Fig. \ref{Fig2}(f)]. This multistep transition is attributed to the metastability of the paraelectric phase [see Fig. \ref{Fig2}(a)]. The maximum critical electric field of 1.2 V/\AA{} is comparable to the electric field strength applied during ferroelectric switching of some 2D ferroelectrics in experiments, such as demonstrated in 2D multiferroic BiFeO$_3$ \cite{Ji2019, Wang2018}. 

\begin{figure}
\centering
\includegraphics*[width=0.48\textwidth]{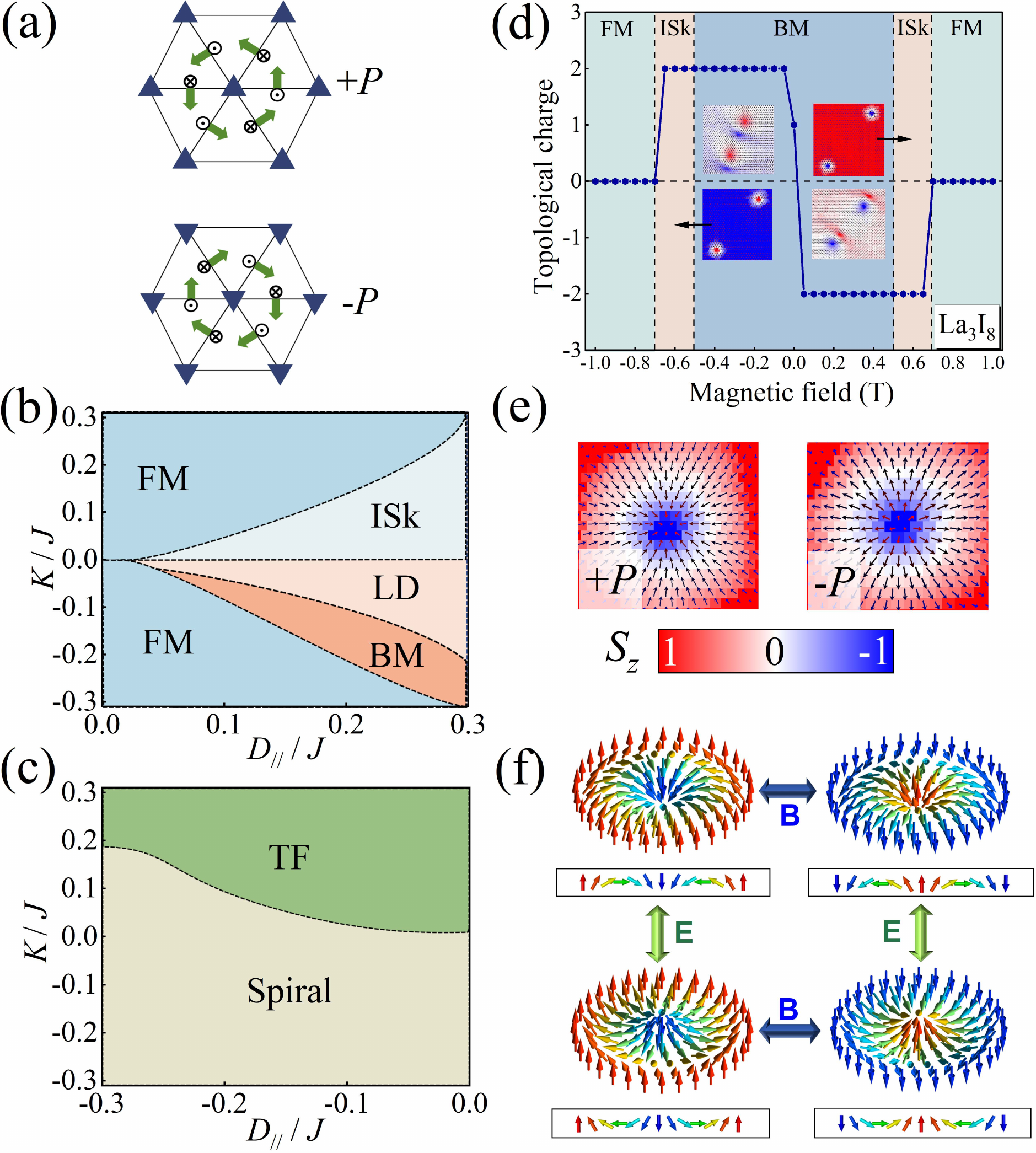}
\caption{\label{Fig3} (a) DM vectors of the nearest-neighbor bonds in two opposite polarization states. Magnetic phase diagrams under (b) ferromagnetic and (c) antiferromagnetic neighbor exchange interaction. The magnetic phases that emerge include ferromagnetic (FM), isolated skyrmion (ISk), labyrinth domain (LD), and bimeron (BM), triangular antiferromagnetic (TF) and spiral spin phases. (d) Transition of magnetic phase and topological charge of La$_3$I$_8$ monolayer with out-of-plane magnetic field. (e) Spin textures of skyrmion appearing in two opposite polarization states. (f) Schematic illustrating the switching of the chirality and topological charge of skyrmions by external electric and magnetic fields, respectively.}
\end{figure}

The breathing mode causes each trimer to carry a magnetic moment, transforming the kagome lattice into a triangular magnetic lattice. In addition, inversion symmetry breaking may give rise to topological spin textures induced by Dzyaloshinskii-Moriya (DM) interactions \cite{Seki2012,Kezsmarki2015,Fert2017,Xu2020,Yu2023}. Switching the breathing mode reverses the DM vectors of adjacent trimers [see Fig. \ref{Fig3}(a)], which is determined by the inversion symmetry correlation between the two opposite polarization states (Fig. S7 \cite{SM}). Figure \ref{Fig3}(b) shows the evolution of the magnetic phase as a function of the DM interaction and the magnetic anisotropy under a ferromagnetic neighbor exchange interaction. With increasing DM interaction, the ferromagnetic phase transforms into an isolated skyrmion or a bimeron phase, depending on the easy-axis or easy-plane anisotropy, respectively [see Fig. \ref{Fig3}(b)], which is consistent with previous observations \cite{Goebel2019,Goebel2021}. The critical value of DM interaction for phase transition increases with the anisotropy constant.  As for the antiferromagnetic exchange interaction, easy-axis anisotropy leads to a spiral spin order, while easy-plane anisotropy energetically prefers the triangular antiferromagnetic structure [see Fig. \ref{Fig3}(c)].

Then the magnetic ground states of all materials with breathing modes were determined by Monte Carlo simulations based on the calculated magnetic parameters (Figs. S8, S9 and Table S1 \cite{SM}). This system exhibits a rich variety of magnetic phases, involving ferromagnetic, triangular antiferromagnetic, spiral spin, and bimeron phases. The emergence of ferromagnetism in some materials such as La$_3$\emph{X}$_8$ (X = Cl, Br) monolayers suggests that this system provides a versatile platform for realizing 2D multiferroics \cite{Zhang2020,Zhang2022,Wang2023a}.

Next, the La$_3$I$_8$ monolayer, exhibiting a topological bimeron phase, is taken as a representative example to demonstrate the modulation effect of an external field. A small out-of-plane magnetic field drives the bimeron phase into the isolated skyrmion phase, which subsequently transforms into the ferromagnetic phase as the magnetic field is further increased [see Fig. \ref{Fig3}(d)]. Applying a magnetic field in the opposite direction induces similar magnetic phase transitions, but with the opposite sign of the topological charge. Similar magnetic phase transitions also occur in other materials (Fig. S10 \cite{SM}). Moreover, in the opposite polarization state, the chirality of the skyrmion is changed due to the reversal of the DM vectors [see Fig. \ref{Fig3}(e)]. Therefore, in this system, the chirality and topological charge of the topological magnetic texture can be controlled by the electric and magnetic fields, respectively, as depicted in Fig. \ref{Fig3}(f). 

\begin{figure}
\centering
\includegraphics*[width=0.48\textwidth]{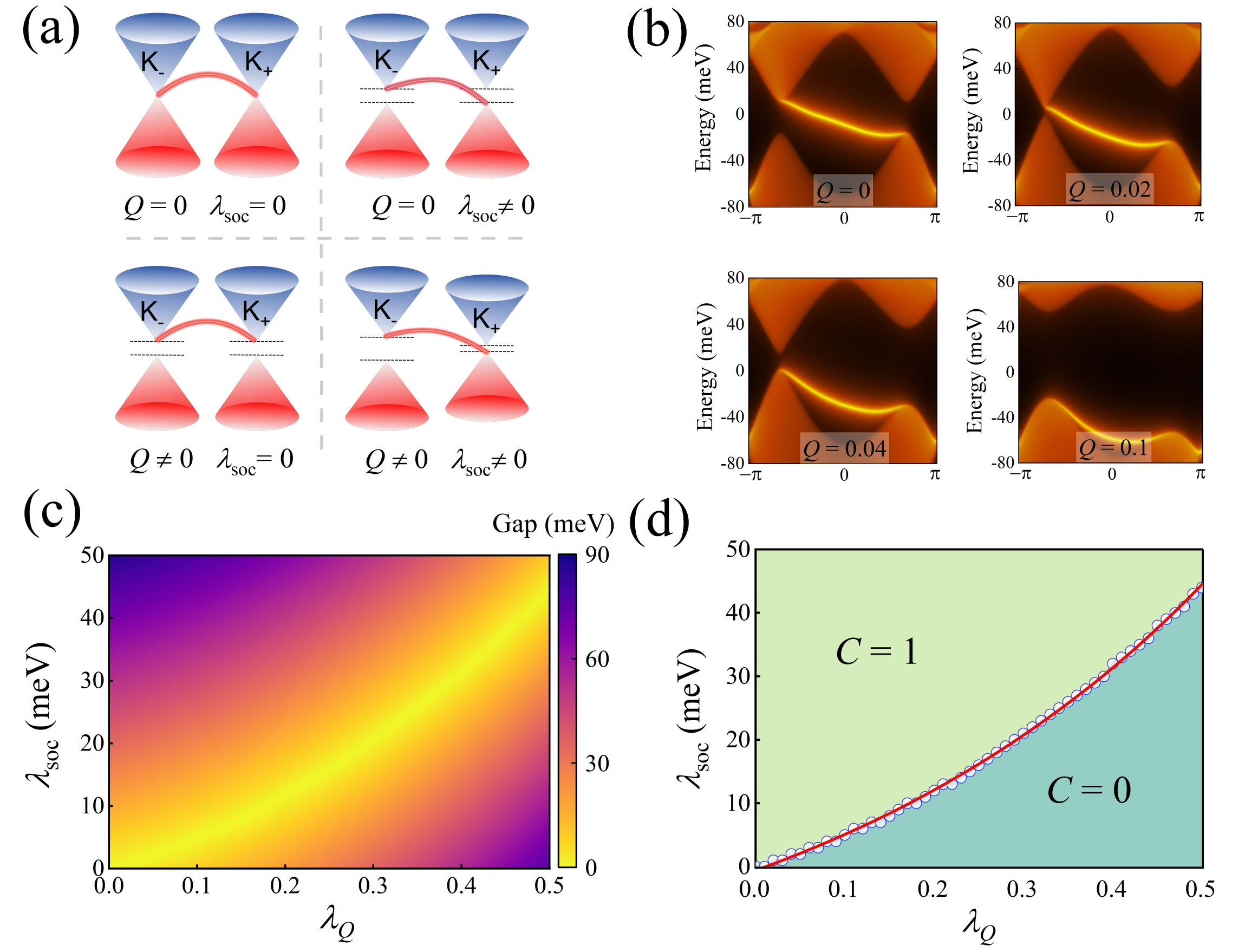}
\caption{\label{Fig4} (a) Schematic of the splitting of the Dirac bands and the resulting edge states in the presence or absence of breathing mode and SOC. (b) Transformation of edge states induced by enhanced breathing mode in La$_3$Cl$_8$ monolayer embedded with K ions. (c) Variation of the band gap with the amplitude of the breathing mode and the SOC constant. The strength of the breathing mode is represented by the relative difference between the nearest-neighbor ($t_1$) and next-nearest-neighbor ($t_2$) transition integrals ($\lambda_{Q} = 1- t_2/t_1$). (d) Chern number as a function of the amplitude of the breathing mode and the SOC constant.}
\end{figure}

In an ideal kagome lattice, the inclusion of spin-orbit coupling (SOC) opens a gap at the Dirac point (K point) \cite{Wang2023}, resulting in a non-trivial topological insulating state with robust gapless edge states [see Fig. \ref{Fig4}(a)]. The breathing mode also opens a band gap at the Dirac point, however, unlike SOC, the induced insulating state is topologically trivial. Despite the broken inversion symmetry, the band gaps at K$_+$ and K$_-$ valleys, are equal due to the protection of vertical mirror symmetry. Including both SOC and breathing mode breaks the valley degeneracy while preserving the non-trivial topological insulating state [see Fig. \ref{Fig4}(a)]. Note that the Fermi level usually does not lie within the band gap at the Dirac points due to the 1/3 filling of the highest occupied \emph{d} orbital. To obtain a realistic topological insulator, the Fermi level of the prototype phase can be raised into the band gap at the Dirac point by inserting alkali metal ions (Fig. S11 and Table S2 \cite{SM}). The optimal site for alkali metal ion absorption is the center of the hexagon, where the magnetic ground state remains unchanged.
 
The band gaps at the Dirac points are significantly affected by the amplitude of the breathing mode. In the absence of SOC, the band gaps at both valleys increase as the breathing mode is enhanced. In contrast, in the presence of SOC, the two band gaps exhibit opposite trends, with one of them closing first and then reopening, signaling a topological phase transition (Fig. S12 \cite{SM}). The topological phase transition caused by the enhanced breathing mode is confirmed by the transition of the edge state [see Fig. \ref{Fig4}(b)]. Tight-binding model calculations show that the critical amplitude of the breathing mode required to close the band gap is proportional to the SOC intensity [see Fig. \ref{Fig4}(c) and Fig. S13 \cite{SM}]. Increasing or decreasing the mode amplitude near the critical value drives the system into topologically trivial and non-trivial insulating phases, respectively [see Fig. \ref{Fig4}(d)]. The critical condition is derived as  $\lambda_{soc} = \frac{\sqrt{3} t_1 \lambda_Q}{2-\lambda_Q}$ (see Supplementary Material \cite{SM}),
which describes the competition effect between SOC and breathing mode in inducing topological phase transition(Fig. S14 \cite{SM}). 

A four-band low-energy effective Hamiltonian model is obtained to reveal the physical mechanism underlying the topological phase transition induced by the breathing mode (see Supplementary Material \cite{SM}), 
\begin{equation*}
H = v\left(\eta \sigma_{x} k_{x}+\sigma_{y} k_{y}\right) \mathbf{1}_{s}+\eta m_{soc} \sigma_{z} s_{z} + m_{b} \sigma_{z} s_{z}+M \mathbf{1}_{\sigma} s_{z},
\end{equation*}
where $\eta =\pm1$ label valley degrees of freedom, $\boldsymbol{\sigma}$ and $\textbf{s}$ are Pauli matrices representing the sublattice and spin degrees of freedom, respectively. The second and third terms represent the effective mass induced by intrinsic SOC and the breathing mode, respectively. These two terms have opposite signs at one of the valleys, causing the total effective mass to gradually decrease to zero as the breathing mode increases and then change its sign. This causes the band gap to close first and then reopen, accompanied by band inversion, resulting in a topological phase transition. Therefore, topological phase transitions can be achieved by modulating the breathing mode through electric field, as realized in La$_3$Cl$_8$.

In conclusion, the $\textit{M}_{3}\textit{X}_{8}$ monolayer system provides a broad platform for studying topological quantum states, topological spin textures and multiferroics related to the breathing mode. The breathing mode mainly occurs in materials containing early 4\emph{d}/5\emph{d} transition metal ions, and materials at both ends of the series exhibit low energy barriers for ferroelectric switching. The breathing mode can be switched in sign or even eliminated by an external electric field. Importantly, the coupling of the breathing mode with magnetic parameters and band topology provides a key degree of freedom for realizing electric-field control of topological spin textures and electronic states. Our discovery presents a paradigm for studying novel topological phenomena associated with the breathing mode in kagome lattice and can inspire the exploration of electric-field control of topological quantum states. 

\begin{acknowledgments} 
This work was financially supported by the National Natural Science Foundation of China (Grants No. 12374097 and No. 11974418), and the Postgraduate Research \& Practice Innovation Program of Jiangsu Province (grant number KYCX24\_2693). Computer resources provided by the High Performance Computing Center of Nanjing University are gratefully acknowledged.
\end{acknowledgments}

%

\end{document}